\documentclass[prl,twocolumn,showpacs]{revtex4}
\usepackage{bm}
\usepackage{amsmath}
\usepackage{amssymb}
\usepackage{graphicx}
\usepackage{amsfonts}
\usepackage{txfonts}
\usepackage{graphicx}

\begin{document}

\title{Helical Quantum States in HgTe Quantum Dots with Inverted Band
Structures}
\author{Kai Chang}
\affiliation{SKLSM, Institute of Semiconductors, Chinese Academy of Sciences, P.O. Box
912, Beijing 100083, China}
\author{Wen-Kai Lou}
\affiliation{SKLSM, Institute of Semiconductors, Chinese Academy of Sciences, P.O. Box
912, Beijing 100083, China}
\date{\today }

\begin{abstract}
We investigate theoretically the electron states in HgTe quantum dots (QDs)
with inverted band structures. In sharp contrast to conventional
semiconductor quantum dots, the quantum states in the gap of HgTe quantum
dot with an inverted band structure are fully spin-polarized, and show
ring-like density distributions near the boundary of the QD and spin-angular
momentum locking. The persistent charge currents and magnetic moments, i.e.,
the Aharonov-Bohm effect, can be observed in such QD structure and oscillate
with increasing magnetic fields. This feature offers us a practical way to
detect these exotic ring-like edge states using the SQUID technique.
\end{abstract}

\maketitle

The exponential growth in the power of computing and information processing
is expected to encounter the quantum limit when the component size reaches
the nanometer scale. Semiconductor quantum dots (QDs) have attracted
intensive attention in the past decades due to its application in electronic
devices \cite{DLoss,Hanson}, e.g., single-electron transistors \cite{Kastner}%
, light-emitting diodes \cite{Park}, diode lasers \cite{Fafard} and solar
cells \cite{Marti}. The electron and hole ground states both localize at the
central part of the QDs, which gives rise to a strong oscillator strength of
the interband optical transition. This is the common feature for various
semiconductor QDs with a positive bandgap, e.g., self-assembled InAs quantum
dots grown by molecular beam epitaxy \cite{Grundmann}, GaAs quantum dots
defined by lithographically patterned gate electrodes \cite{Hayashi}, or by
well-developed etching technique \cite{Kouwenhoven}. Precise control of
quantum states in QDs pave the way to realize highly desirable key functions
in nano-electronics, quantum computing and quantum information processing.
\begin{figure}[tbh]
\includegraphics[width=0.6\columnwidth]{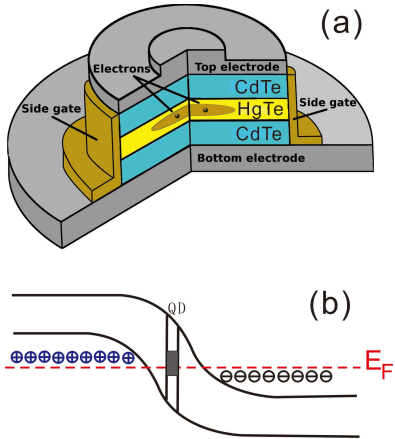}
\caption{(color online) (a) Schematic of a quantum dot formed in a HgTe
heterostructure using etching technique. (b) Schematic of electrical
injection of elecrons and holes into a HgTe QD (the shaded region) embedded
in a p-n junction. }
\end{figure}

By decreasing the bandgap, the band structure can be inverted. For instance,
the bottom of the conduction band near the $\Gamma $ point of the Brillouin
zone in HgTe shows a \textit{p}-like feature and $\Gamma _{8}$ symmetry,
while the top of the valence band shows a \textit{s}-like feature and $%
\Gamma _{6}$ symmetry \cite{XCZhang}. In this case, HgTe becomes insulating
in the bulk with a gap separating the valence and conduction bands, but with
gapless helical edge states that are topologically protected by the time
reversal symmetry, named a topological insulator (TI) \cite{CLKane,BHZ}.
This topological insulator, one of the recent remarkable discoveries, is a
new state of quantum matter \cite{CLKane,XLQi2,FuPRB,XLQi1,KaneReview}. The
quantum spin Hall effect (QSHE) in such two-dimensional TIs was predicted
theoretically\cite{BHZ} and observed experimentally \cite{Konig}. Currently,
the most works focus on searching for new TI materials \cite{NewTI} and
novel transport properties \cite{Hsieh,WYang,BZhou,Sonin}. However, the
electron states in TI nanostructures, e.g., QDs, remains relatively
unexplored.

In this Letter we consider the quantum dots formed in HgTe heterostructure
with an inverted band structure. We find that the fully spin-polarized
quantum states in the gap of the QD show a ring-like density distribution
and spin-angular momentum locking, in sharp contrast to conventional
semiconductor QDs where the electron is localized at the center of the QD.
These states arise from the quantization of the edge states along the
circumference of the QD, and its energies show a approximately linear
dependence on the angular momentum. Importantly, these states are optical
dark states that can be used in quantum information storage. A perpendicular
magnetic field induces the persistent current and magnetic moment which
oscillate with increasing magnetic fields, i.e., the Aharonov-Bohm (AB)
effect. This fundamental quantum effect is never observed in a QD structure.
This effect offer us a possibility for detecting these exotic states using
the SQUID or STM techniques.

We consider a QD formed in a HgTe heterostructure using etching technique as
shown schematically in Fig. 1. \cite{Henz} The low-energy electron states
are described by a four band effective Hamiltonian with a lateral
confinement \cite{BHZ}:%
\begin{equation}
H_{4\times 4}=\left[
\begin{array}{cc}
H(k) & 0 \\
0 & H^{\ast }(-k)%
\end{array}%
\right]  \label{Hamiltonian}
\end{equation}%
where $\mathbf{k}=(k_{x},k_{y})$ is the in-plane momentum of carriers, $%
H(k)=(\varepsilon \left( k\right) )I_{2\times 2}+d^{i}(k)\sigma ^{i}+V\left(
\rho \right) \sigma ^{3}$, $I_{2\times 2}$ is a $2\times 2$ unit matrix, $%
\sigma ^{i}(i=1,2,3)$ the Pauli matrices. $\varepsilon \left( k\right)
=C-D\left( k_{x}^{2}+k_{y}^{2}\right) $, $d^{1}\left( k\right) =Ak_{x}$, $%
d^{2}\left( k\right) =Ak_{y}$, and $d^{3}\left( k\right)
=M(k)=M-B(k_{x}^{2}+k_{y}^{2})$. The parameters $A,B,C,D,M$ depend on the
thickness of HgTe quantum well. The Hamiltonian is obtained by reducing the
eight-band Kane model to the reduced Hilbert space $\left \vert e\uparrow
\right \rangle $, $\left \vert hh\uparrow \right \rangle $, $\left \vert
e\downarrow \right \rangle $, and $\left \vert hh\downarrow \right \rangle $%
. Notice that the Hamiltonian is block-diagonal and possesses the time
reversal symmetry for the upper and lower $2\times 2$\ blocks. $M$ is an
important parameter that can be used to describe the band insulator with a
positive gap ($M>0$) and topological insulator with a negative gap ($M<0$)
cases. The confining potential $V\left( \rho \right) $ of QD can be
simulated by a hard-wall potential: $V\left( \rho \right) =0$, for $\rho <R$%
; $\infty $ otherwise; $R$ is the radius of the hard-wall disk. Other
confining potentials, e.g., parabolic confinement, show the essentially same
energy spectrum, we omit it here for the brevity. The eigenstates and
eigenenergies can be obtained numerically by expanding the wave function $%
\Psi _{i}=\sum_{n,m}C_{n,m}^{\left( i\right) }\varphi _{n,m}$ in the terms
of the Bessel basis for the hard-wall disk, where the index $i$ corresponds
to the different spin projection $S_{z}=$ $\pm 1/2,\pm 3/2$. $%
C_{n,m}^{\left( i\right) }$ is the expanding coefficient. For a hard-wall
disk, the basis function $\varphi _{nm}$ can be expressed as $\varphi _{nm}=%
\mathbb{N}_{C}J_{m}\left( k_{n}^{m}\rho /R\right) e^{im\varphi }$, where $%
k_{n}^{m}$ is the $n$-th zero point of the first kind of the Cylinder Bessel
functions $J_{m}\left( x\right) $, $\mathbb{N}_{C}=1/\left[ \sqrt{\pi }%
RJ_{m+1}\left( k_{n}^{m}\right) \right] $. $m=0,\pm 1,\pm 2...$ is the
quantum number of the angular momentum. Because the Hamiltonian has
cylindrical rotation symmetry, the total angular momentum along the z
direction is in conservation, $\left[ \hat{J}_{z},\hat{H}_{0}\right] =0$.
Where $\hat{J}_{z}=\hat{L}_{z}+\hat{S}_{z}$, $\ L_{z}$ is the orbit
azimuthal angular momentum and $\hat{S}_{z}$ is the total spin $\hat{S}$
projection onto $z$ direction. Therefore the total azimuthal angular
momentum quantum number $j$, eigenvalue of $J_{z}$, is good quantum number.
Also we notice that the off-diagonal elements in the Hamiltonian (1) only
couple the basis functions with the angular momenta $e^{im\varphi }$ and $%
e^{i(m+1)\varphi }$. In the presence of an external perpendicular magnetic
field, the canonical momentum is $\mathbf{P}=\mathbf{p}+e\mathbf{A}$. $%
\mathbf{A}=\mathbf{B}\times \mathbf{r}/2$ is the vector potential adopting
the symmetric gauge. The number of the eigenstates used in the expansion is
chosen to ensure the convergence of the calculated\ energies of the quantum
states within and nearby the bulk gap.

For the edge states, we can obtain the analytical expression of the energy
and wavefunction. Using the trial wavefunction\cite{BZhou} $\psi =e^{\lambda
\left( \rho -R\right) }(e^{im\varphi },e^{i(m+1)\varphi })^{T}$ in the polar
coordinate. the virational parameter $\lambda $ can be given from the
secular equations for the upper block of the Hamiltonian (1),%
\begin{equation}
\left[ \lambda ^{2}-(m/R)^{2}\right] ^{2}-\left( a+b\right) \left[ \lambda
^{2}-(m/R)^{2}\right] -c=0,
\end{equation}%
where $a=-2\left[ D\left( E-C\right) +BM\right] /B_{+}B_{-}$, $%
b=A^{2}/B_{+}B_{-}$, $F=(a+b)/2$, $c=-(M^{2}-E^{2}-C^{2}+2CE)/B_{+}B_{-}$.
To obtain the above equation, we take $\rho \approx R$ because the edge
states are localized at the boundary of the QD. The variation parameter and
the energy can be obtained as%
\begin{eqnarray}
\lambda _{1,2} &=&\sqrt{(\frac{m}{R})^{2}+F\pm \sqrt{F^{2}+\frac{%
E^{2}-M^{2}+C^{2}-2CE}{B_{+}B_{-}}}}, \\
E &=&\frac{BC-DM}{B}\pm A\sqrt{1-\left( \frac{D}{B}\right) ^{2}}m/R.
\end{eqnarray}%
\begin{figure}[tbh]
\includegraphics[width=1.0\columnwidth]{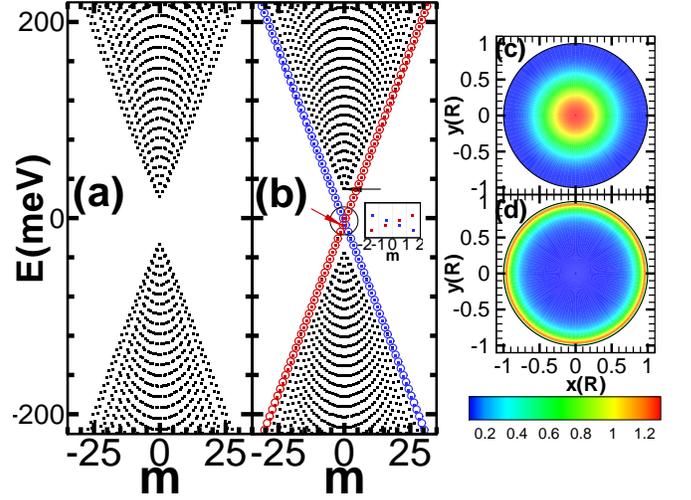}
\caption{(color online) The energy spectra of the HgTe QDs in band insulator
($M>0$) and topological insulator ($M<0$) cases as a function of the angular
momentum $m$. (a) is in the band insulator phase ($M>0$); (b) is in the
topological insulator phase ($M<0$). The new quantum states are denoted by
the red (rotating clockwise) and blue (rotating counter-clockwise) squares.
The red and blue circles are plotted from the analytical solutions Eq. (3).
The inset is the amplification of the circle region near the Dirac point.
(c) The density distributions of the lowest conduction band electron states
[marked by the black arrow in Fig. 2(b)] in the QD; (d) The same as (c), but
for the RES [marked by the red arrow in Fig. 2(b)] near the Dirac point. The
QD radius $R=55nm$, $M=-30meV$.}
\end{figure}

We show the energy spectra of a HgTe QD with a hard-wall confining potential
as a function of the angular momentum $m=<L_{z}>$ in Fig. 2. The relevant
parameters used in our calculation are obtained from Ref. \cite{BHZ}. In
Fig. 2, we consider two different cases: the band insulator ($M>0$) and the
topological insulator ($M<0$) cases. Comparing Fig. 2(a) with Fig. 2(b), one
can see that the quantum dot spectrum is gapped for the band insulator QD
case ($M>0$), which is similar with that of the conventional semiconductor
QDs. But for the topological insulator QD case ($M<0$), the new quantum
states emerge in the whole energy spectrum even in the gap. These states
show a linear energy dispersion, i.e., the Dirac spectrum, against the
angular momentum $m$ in the bulk gap. The energy spectrum of these states
(the red and blue squares and circles in Fig. 2(b)) shows symmetric respect
to the angular momentum $m$ due to the time reversal symmetry. In order to
understand the physical origin of these states, we plot the density
distribution of the quantum states appearing within the bulk gap in Figs.
2(c) and 2(d). Figure 2(c) shows the density distributions of the lowest
conduction band electron states [marked by the black arrow in Fig. 2(b)],
which peak at the center of the QDs as in conventional semiconductor quantum
dots. Figure 2(d) describes the density distribution of the new quantum
states [marked by the red arrow in Fig. 2(b)] nearby the Dirac point.
Surprisingly, one can see the state localizes at the edge of the hard-wall
QD, and shows the ring-like density distribution. We would stress that all
these new quantum states carry approximately half-integer angular momentums $%
m=I\pm 1/2$, where the integer $I=0,\pm 1,\pm 2,...$[see the inset of Fig.
2(b)], and show the homocentric ring-like density distributions. We name
these exotic quantum states as the ring-like edge states (RESs) in this
paper. We calculate the dependence of the energies of the bulk QD states and
the RESs on the size of the QD in Fig. 3(a). For the quantum states arising
from the quantized electron bulk states localized at the center of the QD,
the energy spectrum shows a linear dependence on the inverse of the area of
the QD [see the inset of Fig. 3(a)], i.e., $1/R^{2}$. Interestingly, the
energies of the RESs display a linear dependence on the inverse of the
circumference of the QD, i.e., $1/R$. This feature indicates clearly that
these exotic quantum states come from the quantization of the edge states in
two-dimensional topological insulators along the circumference of the
hard-wall disk. The RESs are equally spaced in the energy spectrum [see Fig.
2(b)] because of the new quantization rule $m=I\pm 1/2$ and show the linear
energy dispersion against the angular momentum $m$. This characteristic
means that electrons in the RESs behave like massless Dirac fermions. When
this massless Dirac fermion is confined in a disk, its lowest energy modes
should be the whispering gallery mode, similar to a photon confined in a
cylinder cavity. This gives us an intuitive picture to understand the origin
of these exotic ring-like quantum states in QDs. From Fig. 3(b) one can see
that the energy level spacing $\Delta E=E_{m}-E_{m-1}$ of the RESs (the gap
of the bulk states $E_{g}=E_{m=0}^{c}-E_{m=0}^{v}$, the indexes $c$, $v$
denote the conduction and valence bands, respectively) in the QD increase
linearly as the inverse of the radius $1/R$ (area $1/R^{2}$) of the QD
increases. These interesting features of the edge states mentioned above can
also be seen from the analytical expression of the edge state energy (see
Eq. (4) and the circles in Figs. 3(a) and 3(b)). The slight difference
between the numerical and analytical results arises from the approximation $%
\rho \approx R$ adopted in our derivation.

\begin{figure}[tbh]
\includegraphics[width=1.\columnwidth]{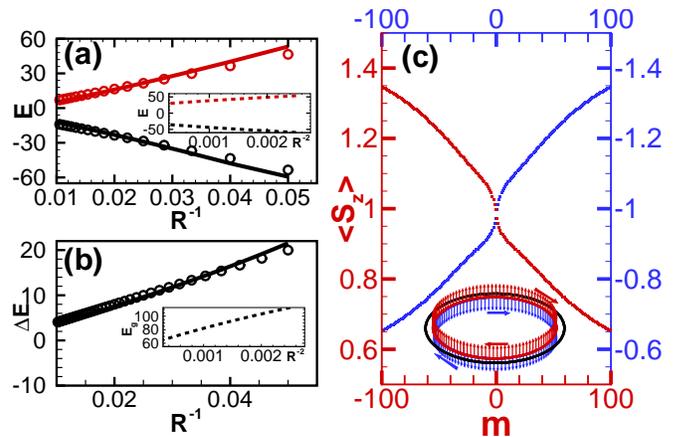}
\caption{(color online) (a) The energies of the up (the red line) and down
(the black line) branches of the RESs \textit{vs} the inverse of the radius
of the QD $1/R$. The inset shows the dependence of the energies of the
electron states in the bulk conduction (the red dashed line) and valence
(the black dashed line) bands on the inverse square of the radius of the QD $%
1/R^{2}$, respectively. (b) The energy level spacing $\Delta E$ of the RESs
\textit{vs} the $1/R$, the inset shows the bulk energy gap of the QD \textit{%
vs} $1/R^{2}$. The circles in (a) and (b) are obtained from the analytical
expression of the edge states. (c) The spin projection $<S_{z}>$ for the
ring-like edge states with the different angular momentum $m$. The inset is
the schematic of the spin orientation of the RESs in the QD. $R=55nm$, $%
M=-30meV$. }
\end{figure}

Interestingly, these ring-like states are fully spin-polarized and show a
spin-angular momentum locking. In Fig. 3(c), there are two kinds of these
ring-like quantum states in which spin-up electrons rotate clockwise and
spin-down electron rotate counterclockwise [see Fig. 3(c)]. This opposite
spin orientation is the consequence of the time reversal symmetry in the
four-band Hamiltonian $(1)$. Note that the spin projection of edge states $%
<S_{z}>$ is not a good quantum number, and varies significantly with
increasing the angular moment $m$. For the edge states rotating clockwise
(the spin-up branch) and counterclockwise (the spin-down branch), the spin
orientation $<S_{z}>$ varies approximately from $0.6$ to $1.4$, and $-0.6$
to $-1.4$, respectively, which indicates that the mixing between the
electron and the hole states changes with the angular momentum and therefore
destroys the spin conservation. The spin orientation $<S_{z}>$ is symmetric
respect to the angular momentum $<m>$, i.e., rotating clockwise and
counterclockwise. Interestingly, this degeneracy $E_{m}=E_{-m}$ is lifted by
a weak magnetic field, thus one can get a fully spin-polarized electron
states by controlling the number of electrons in such a QD at low
temperatures.

\begin{figure}[tbh]
\includegraphics[width=1.\columnwidth]{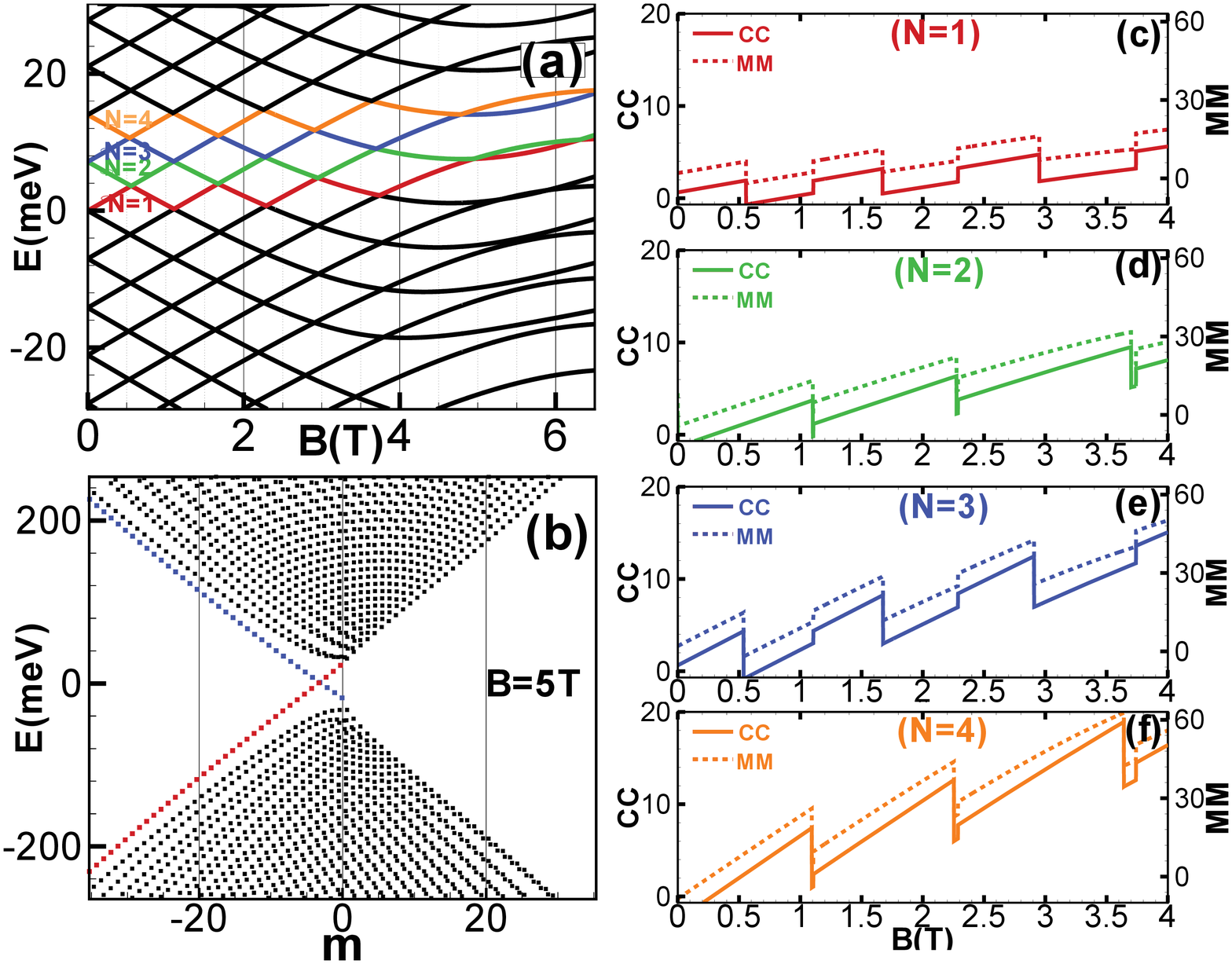}
\caption{(color online) (a) The magnetic level fan in a HgTe QD with a
inverted band structure. (b) The energy spectrum in the presence of magnetic
fields for the TI QD. (c)-(f) The persistent charge current (CC) (the solid
line) and magnetic moment (MM) (the dashed line) in a TI QD against the
magnetic fields for different electron number $N$. The persistent CC is in
units of $2NE_{0}/\Phi _{0}$ ($2N\ast 2nA$ for the hard wall disk, where $%
E_{0}=\hbar ^{2}/[2m^{\ast }(R)^{2}]$. $R=55nm$, $M=-30meV$. The magnetic
moment is in units of $N\protect \mu _{B}$, where $\protect \mu _{B}=e\hbar
/(2m_{e})$. The red, green, blue and orange curves indicates different
electrons $N=1,2,3,4$ filled in the edge states (see the panel (a)). }
\label{fig:fig4}
\end{figure}

Next we turn to discuss the optical property of these edge states in QDs.
The absorption spectra of the QD $\alpha (\hbar \omega )=$ $\frac{\pi e^{2}}{%
m_{0}^{2}\epsilon _{0}cn\omega V}\sum_{i,f}\left \vert \overrightarrow{%
\varepsilon }\cdot P_{if}\right \vert ^{2}\delta (E_{f}-E_{i}-\hbar \omega )$%
, where $n$ is the refractive index, $c$ is the speed of light in vacuum, $%
\epsilon _{0}$ is the permittivity of vacuum, $m_{0}$ is the free-electron
mass, and $\overrightarrow{\varepsilon }$ is the polarization vector of the
incident light. We are only interested in the vertical optical transition
between the up and down branches of the edge states near the Dirac point,
i.e., within in the bulk gap. The momentum matrix $\mathbf{P}\propto
\partial H_{4\times 4}/\partial \mathbf{k}$ is block diagonal and will not
couple the up and down branches of the edge states belonging to the spin-up
and spin-down families, therefore the vertical circular transition between
the up and down branches is forbidden because of the conservation of the
angular moment. It makes it hardly possible to detect these ring-like edge
states using optical techniques, e.g., photoluminescence. It means that
these RESs are optically inactive within the electric dipole approximation,
i.e., the dark states. However, the RESs also provides us a possibility to
storage quantum information in these dark states, e.g., a electron-hole pair
in the RESs formed by electric injection [see Fig. 1(b)].

Finally we study the effect of magnetic field on these RESs. We consider an
external magnetic field applied perpendicularly to HgTe QD plane. Fig. 4(a)
shows the magnetic levels in a HgTe QD ($M<0$)\ including the bulk and edge
states. The energy spectra display many crossing points in the bulk gap
region. This is because an electron in these edge states shows a ring-like
density distribution, and the energy spectrum is similar with that of a
quantum ring. The energy spectra become no longer symmetric with respect to
the angular moment $<m>$ because the magnetic field breaks the time-reversal
symmetry [see Fig. 4(b)]. The Dirac point is shifted left, and the edge
states rotating clockwise (counterclockwise) is strongly squeezed
(separated), since the Lorentz force induced by the magnetic field
suppresses (enhances) electron rotating clockwise (counterclockwise). At
high magnetic field ($R>>l_{B}$), the edge states rotating clockwise are
pushed into the bulk states and can not be seen in the spectra as shown in
Fig. 4(b).

It is well known that a magnetic field can induce a persistent current in
mesoscopic rings and even type-II quantum dots, i.e., Aharonov-Bohm effect%
\cite{Sellers,BardarsonAB,ZhangAB}, which oscillates with increasing
magnetic fields. Such quantum mechanical phenomenon is never observed in
type-I QD systems. Interestingly, this phenomenon can be found in this
topological insulator QD system, because the unique density distributions of
the RESs show a ring-like behavior and almost localized at the boundary of
the hard-wall disk. The persistent current for a single TI QD can be
obtained by $I=\int \hat{\jmath}_{c}\left( r\right) d\overrightarrow{r}$,
where the current density operator $\hat{\jmath}_{c}\left( r\right) =\left[
\hat{\rho}\hat{\upsilon}+\hat{\upsilon}\hat{\rho}\right] /2$, and the
magnetic moment $\vec{M}$ generated by the persistent current is given by: $%
\vec{M}=\int \vec{r}\times \hat{\jmath}_{c}\left( r\right) d\vec{r}/2$,
where $\hat{\rho}$ is the electron density operator, $\hat{\upsilon}$ is the
electron group velocity operator along the tangential direction. From Figs.
4(c)-4(f), one can see that the magnitude of the magnetic moment induced by
the persistent current depends on the electron numbers in the QD and ranges
from $20\sim 200\mu _{B}$ , which is beyond the limit of resolution and
sensitivity of near-field SQUID magnetometry (around $1\sim 10\mu _{B}$)
\cite{nanoSQUID}. The persistent charge current and the magnetic moment are
a periodic function of magnetic fields, exhibiting many linear segments with
a fixed slope ratio. The periodicity of the persistent charge current for
different electron number $N$ is almost the same. This feature can be
understood from the crossing points in the magnetic energy level spectrum in
Fig. 4(a). When the magnetic field sweeps across the points, the persistent
current and magnetic moment oscillate. The oscillation of the magnetic
moment provides us a possibility to detect these exotic quantum states in
HgTe QDs using near-field SQUID technique at low temperature \cite%
{nanoSQUID,SQUID}.

In summary, we study the exotic quantum states in quantum dots formed in
HgTe heterostructure with inverted band structures. As a consequence of the
quantization of the edge states, these states appear in the bulk gap of the
QDs and show ring-like density distributions. The RESs are optically
inactive since the vertical optical transition between these states is
forbidden. This feature is in sharp contrast to that of the conventional
semiconductor quantum dots with normal band structures. The oscillating
persistent currents and magnetic moments can be found as the perpendicular
magnetic fields increase, which make it possible to detect these quantum
states utilizing the nano-SQUID technique. This work shed a new light on
constructing topological insulator-based nano-electronic devices.

\begin{acknowledgments}
This work was partly supported by the NSFC Grant Nos. 60525405 and 10874175,
and the financial support from the CAS and MOST.
\end{acknowledgments}

\end{document}